\documentclass[11pt]{article}
\usepackage{englwin}
\usepackage{ihep}
\usepackage{ipsizen}

\usepackage{amsmath}
\usepackage{author}
\usepackage[colorlinks]{hyperref}

\begin{document}
\begin{titlepage}

\prepnum{2004--36}{}

\author{\Large  V.V.~Ezhela,~Yu.V.~Kuyanov,~V.N.~Larin,~A.S.~Siver}

\title{
\Large \bf {T\uppercase{he inconstancy \\of~the~fundamental~physical~constants:
computational~status}}}


\end{titlepage}

\begin{abstractpage}[539.1.01]
\numref{11}
\engabs
{Ezhela V.V., Kuyanov Yu.V., Larin V.N., Siver A.S.}
{The Inconstancy of the Fundamental Physical Constants: Computational Status}

It is argued that the CODATA recommended values of the fundamental
physical constants  could not be used as the reference data in
searching the hypothetical space-time variations of the
fundamental physical constants.

It is shown that the CODATA data permanently suffers a loss of
 self-consistency of the released data due to unjustified
 over-rounding of their estimates.

The simple estimates of the critical numbers of decimal digits
that should be saved in the independently rounded correlation
coefficients, the average values and uncertainties to save the
self-consistency is obtained.

The set of high level quality requirements to the computerized
presentation of the numerical data on the jointly measured or estimated
physical values are formulated.

It is argued (once again) that the common standard for
presentation of the numerical values of correlated quantities in
publications and sites is urgently needed.

\rusabs
{Ежела В.В., Куянов Ю.В., Ларин В.Н., Сивер А.С.}
{Непостоянство фундаментальных физических постоянных: вычислительный статус}

Приведены свидетельства того, что рекомендуемые CODATA значения
фундаментальных физических постоянных непригодны для проверки гипотезы о возможном различии значений фундаментальных постоянных в разных областях во времени и пространстве.

Показано, что публикуемые CODATA таблицы значений как на бумажных
носителях, так и в электронном виде, испорчены некорректным
округлением численных представлений средних значений, стандартных
отклонений и коэффициентов корреляций.

Представлены простые оценки точностей корректного
представления округленных средних значений, стандартных отклонений и
коэффициентов корреляций. Эти оценки можно использовать для
контроля корректности и согласованности значений фундаментальных
физических постоянных.

Сформулированы предложения по общим требованиям к качеству представления числовых данных
о совместно измеренных или оцененных физических величинах: их  средних значений, стандартных
отклонений и коэффициентов корреляций в публикациях, справочниках
и на сайтах.

\end{abstractpage}

\noindent
\boldmath
\vspace*{-1.5cm}
\section{Motivation}
The possible space and time variations of the fundamental physical
constants (FPC) continuously attract much attention of different
investigators since the time when Dirac has invented the idea.
Following a recent review of J.P.~Uzan \cite{Uzan:2002vq}, a
general strategy for searches of the variability can be  outlined
as follows:

\begin{itemize}
\item The hypothesis of constancy of the FPC can and must be
checked experimentally.

\item It only make sense to consider the variations of
dimensionless combinations (ratios) of the fundamental constants.

\item If the FPC vary, they most probably vary jointly and slowly.
This means that to notice FPC variations we should:
\begin{itemize}
\item  select several well separated space-time regions;

\item measure/estimate as precise as possible physics observables
expressed in terms of the FPC, that refer to the same space-time
region;

\item compare values of constants in the different space-time
regions, but extracted from the ``space-time region dependent"
observables with the same current FPC evaluation and adjustment
methods.
\end{itemize}
\end{itemize}

Let $V_{X,i}$ denotes the set of  FPC related random variables to
be  estimated and adjusted by the method of least squares (for
example) on the experimental data at space-time region $X$. This
means that after the successful adjustment we will have in the
parametric $V$-space the vector of average values $\langle V_{X,i}
\rangle$ and the corresponding covariance matrix
$Cov(U_{X,i},U_{X,j})$, characterizing the interior of the
``scatter ellipsoid'' centered at the end of the vector of
averages
\begin{equation}
\sum_{ij} (V_{X,i} -\langle V_{X,i} \rangle)\cdot [Cov(U_{X,i},U_{X,j})]^{-1}
\cdot(V_{X,j}-\langle V_{X,j} \rangle) < 1.
\end{equation}
The same ellipsoid can be represented with the help of correlation
matrix $C_{ij}(X) =$ $Cor(U_{X,i},$ $U_{X,j})$ and standard deviations
$U_{X,i} = \sqrt{Cov(U_{X,i},U_{X,i})}$ of $V_{X,i}$.
\begin{equation}
\sum_{ij}{\frac {V_{X,i} - \langle V_{X,i} \rangle} {U_{X,i}}}
\cdot [C_{ij}(X))]^{-1}\cdot {\frac {V_{X,j}- \langle V_{X,j} \rangle} {U_{X,j}}}< 1.
\label{corellipsoid}
\end{equation}
To see the space-time variability we should see the well separation of the
scatter ellipsoids in the $V$-space. Let us say that vector $V$
deviates from the scatter ellipsoid obtained for $X$ space-time region by
$ R_X(V,\langle V_X\rangle)$ standard deviation if
\begin{equation}
\sum_{ij} (V_i -\langle V_{X,i} \rangle)\cdot [Cov(V_{X,i},V_{X,j})]^{-1}
\cdot(V_j-\langle V_{X,j} \rangle) = R^2_X(V,\langle V_X\rangle).
\end{equation}
Then it is easy to see that the scatter ellipsoids obtained for
the $X$ and $Y$ regions will be well separated if in the whole $V$
space we will have
\begin{equation}
R^2_X(V,\langle V_X\rangle)+ R^2_Y(V,\langle V_Y\rangle) > 2,
\end{equation}
that means that the scatter ellipsoids do not intersect. Hence, to
be able to notice the variability we should have both: accurately
estimated average values and corresponding scatter ellipsoid for
every space-time region where we estimate the FPC. It is the
delicate problem as we will show further.

 The only and the best known well elaborated procedures to evaluate and adjust
 fundamental physical constants are implemented at the NIST Physics Laboratory \cite{NIST}.
 The set of FPC periodically adjusted at NIST is
recommended by CODATA as the reference source of the FPC for
scientific applications and technology. In any attempt to notice
the space-time variability of the FPC one cannot avoid the  CODATA
recommended values, deemed in the physics community as the one of
the best known set of FPC adjusted in the space-time region where
we are.
But unfortunately it is impossible. Simply because we never had the
set of the recommended FPC correct enough for the testing their
space-time variability.
To show this let us select subsample of the dimensionless
FPC from the CODATA-2002 recommended set \cite{CODATA:02}, say the set: \\[-2mm]

\noindent
\begin{tabular}{|l|c|l|l|}
\hline
Standard FPC name & Symbol& Value (2002)& Uncertainty\\ \hline \hline
 fine-structure constant&$\alpha$&7.297 352 568e-3&0.000 000 024e-3 \\
 \hline
 electron-muon mass ratio&$m_e/m_{\mu}$& 4.836 331 67e-3&0.000 000 13e-3\\
\hline
electron-proton mass ratio&$m_e/m_p$&5.446 170 2173e-4&0.000 000 0025e-4\\
\hline
electron-deuterium mass ratio&$m_e/m_d$&2.724 437 1095e-4&0.000 000 0013e-4\\
\hline
electron-proton magn. moment ratio&$\mu_e/\mu_{p}$&-658.210 6862&
0.000 0066\\
\hline
muon-proton magn. moment ratio&$\mu_{\mu}/\mu_p$&-3.183 345 118&0.000 000 089\\
\hline
proton $g$ factor & $g_p=2\mu_p/\mu_n$& 5.585 694 701& 0.000 000 056\\
\hline
\end{tabular}\\ [2mm]

\noindent
The corresponding CODATA-2002 correlation matrix is as follows:
\begin{center}
\begin{tabular}{|l||c|c|c|c|c|c|c|}
\hline
& & & & & & & \\[-4mm]
 $Cor(2002)$         & $\alpha$&$m_e/m_{\mu}$&$m_e/m_p$&$m_e/m_d$&$\mu_e/\mu_{p}$&$\mu_{\mu}/\mu_{p}$&
 $g_p=2\mu_p/\mu_n $ \\
\hline \hline
$\alpha$             & 1.000 & -0.247 & 0.000& 0.000&  -0.003&  0.230& -0.002\\
 \hline
$m_e/m_{\mu}$        & -0.247&1.000   & 0.004& 0.004&   0.008& -0.934&  0.008\\
\hline
$m_e/m_p$            & 0.000 & 0.004  & 1.000& 0.894&   0.000& -0.004& -0.046\\
\hline
$m_e/m_d$            & 0.000 & 0.004  & 0.894& 1.000&   0.000&  0.012& -0.041\\
\hline
$\mu_e/\mu_{p}$      & -0.003& 0.008  & 0.000& 0.000&   1.000& -0.008&  0.999\\
\hline
$\mu_{\mu}/\mu_{p}$  & 0.230 &-0.934  &-0.004& 0.012&  -0.008&  1.000&  0.350\\
\hline
$g_p=2\mu_p/\mu_n $  &-0.002 & 0.008  &-0.046& 0.041&   0.999&  0.350&  1.000\\
\hline
\end{tabular}
\end{center}
This matrix is non-positive definite matrix (it has one negative eigenvalue
  $= -0.000293338$).

This means that we have no scatter ellipsoid, the corresponding
``scatter region'' is unbounded and the comparison with any other
evaluations is senseless. This confusion might be due to misprints
in the resource database as of 2002, but this is not the case. The
same situation with non-positive definite correlation matrices is present
in all releases of the FPC produced by NIST and
approved/recommended by CODATA. Further examples of the wrong
subsamples of the CODATA recommended FPC see in the Table
\ref{tab1}, where we compare data from the last three releases
(V.3.0, V.3.2, V.4.0). The other examples presented also in our
previous papers \cite{IMS:03,ASS-VVE:03} on this subject.\\[-5mm]

\begin{table}[h]
\label{tab1}
\begin{center}
\caption{Comparison of the selected CODATA:1986, CODATA:1998, and CODATA:2002 recommended values for the triads of quantities: averages, uncertainties, correlations.}
\small
\vspace*{1mm}
\begin{tabular}{|l|c|l|ccc|}
\hline
 & & & \multicolumn{3}{|c|}{}\\[-3mm]
{\bf \large CODATA:1986} & Symbol [units] & Value (uncertainty)$\times$scale& \multicolumn{3}{|c|}{Correlations}\\[1mm]
\hline
& & & & & \\[-4mm]
Elementary charge & $e \, \,\quad \quad [{\rm C}]$ & $ 1.602\, 177\, 33(49)\times10^{-19}$
& $e$ & $h$ & $m_e$ \\ \cline{4-6}
& & & & & \\[-4mm]
 Plank constant & $h \quad \quad [{\rm J} \, {\rm s}]$ &
 $6.626\, 075\, 5(40) \times 10^{-34}$ &$\phantom{-}$0.997 &  & \\
Electron mass & $m_e \, \quad [{\rm kg}]$ & $9.109\, 389\, 7(54) \times 10^{-31}$
&$\phantom{-}$0.975 &$\phantom{-}$0.989 & \\
1/$\alpha(0)$ & $\alpha(0)^{-1} \, \, \quad $&$ 137.035 \, 989\, 5(61)$
&$-$0.226 &$-$0.154 &$-$0.005 \\ [1mm]
\hline \hline
 & & & \multicolumn{3}{|c|}{}\\[-3mm]
{\bf \large CODATA:1998 } & Symbol [units] & Value (uncertainty)$\times$scale& \multicolumn{3}{|c|}{Correlations}\\[1mm]
\hline
& & & & & \\[-4mm]
Elementary charge & $e \, \, \quad \quad [{\rm C}]$ & $ 1.602\, 176\, 462(63)\times10^{-19}$
& $e$ & $h$ & $m_e$ \\ \cline{4-6}
& & & & & \\[-4mm]
 Plank constant & $h \quad \quad [{\rm J} \, {\rm s}]$ &
 $6.626\, 068\, 76(52) \times 10^{-34}$ &$\phantom{-}$0.999 &  & \\
Electron mass & $m_e \, \quad [{\rm kg}]$ & $9.109\, 381\, 88(72) \times 10^{-31}$
&$\phantom{-}$0.990 &$\phantom{-}$0.996 & \\
1/$\alpha(0)$ & $\alpha(0)^{-1} \, \quad$ & $137.035 \, 999\, 76(50)$
&${-}$0.049 &${-}$0.002&$\phantom{-}$0.092 \\ [1mm]
\hline \hline
 & & & \multicolumn{3}{|c|}{}\\[-3mm]
{\bf \large CODATA:2002 } & Symbol [units] & Value (uncertainty)$\times$scale& \multicolumn{3}{|c|}{Correlations}\\[1mm]
\hline
& & & & & \\[-3mm]
Elementary charge & $e \, \, \quad \quad [{\rm C}]$ & $ 1.602\, 176\, 53(14)\times10^{-19}$
& $e$ & $h$ & $m_e$ \\ \cline{4-6}
& & & & & \\[-3mm]
 Plank constant & $h \quad \quad [{\rm J} \, {\rm s}]$ &
 $6.626\, 0693(11) \times 10^{-34}$ &$\phantom{-}$1.000 &  & \\
Electron mass & $m_e \, \quad [{\rm kg}]$ & $9.109\, 3826(16) \times 10^{-31}$
&$\phantom{-}$0.998 &$\phantom{-}$0.999 & \\
1/$\alpha(0)$ & $\alpha(0)^{-1} \, \quad$ & $137.035 \, 999\, 11(46)$
&$-$0.029 &$-$0.010 &$\phantom{-}$0.029 \\ [1mm]
\hline
\end{tabular}
\end{center}
\end{table}

\normalsize
\noindent
The eigenvalues of these correlation sub-matrices are as follows:\\

$ CODATA:1986 \quad \{2.99891,\, 1.00084,\, \phantom{-}0.000420779,\, -0.000172106\};$

$ CODATA:1998 \quad \{2.99029,\, 1.01003,\,-0.000441572,\, \phantom{-}0.00012358\};$

$ CODATA:2002 \quad \{2.99802,\, 1.00173,\, \phantom{-}0.000434393,\,-0.000183906\}.$\\

\noindent Definitely something is wrong with the NIST
evaluation/adjustment/presentation procedures. We suspect that the
origin of these permanent confusions is the unjustified
independent rounding of the output interrelated quantities: vector
of constant estimates, their standard deviations(uncertainties)
and their correlations.

Superficial independent rounding may lead to catastrophic changes in the
connection of averages, standard uncertainties and the scatter
ellipsoid: the rounded average values may get out of the
``etalon'' scatter ellipsoid obtained after rounding the
correlation matrix. The ``scatter region'' may turn to become
hyperboloid. From the other hand any numerical calculation is
performed with rounding or truncating decimal numbers.

To preserve the general properties of the FPC data structure, a
special quality assurance procedures should be developed and
applied. In the next section we collect the high level
requirements to the set of FPC needed to guarantee the safe
and correct usage of this key informational resource.

\section{High level requirements to the set of adjusted FPC}

Let us introduce a few special notations and definitions for different sets of FPC
to simplify formulation and discussions of the  requirements.\\

$V^B$ or {\bf ``basic FPC''} is the set of constants that participated in the fits to the experimental data via observational equations. \\

$V^D$ or {\bf ``derived FPC''} is the set of constants and units
conversion factors that are known to be function dependent on
basic constants. Symbolically  $V^D = F(V^B)$ and they are
evaluated on the basis of the $V^B$ with the proper propagation of
the uncertainties with the sufficient accuracy to guarantee
positive semidefinitness of the derived covariance
martix\footnote{ By definition the covariance (correlation) matrix
for the jointly measured or estimated quantities is the positive
semidefinite matrix, moreover if adjustment is performed by the
least squares method the covariance (correlation)
matrix if presented with the etalon accuracy should be positive definite for the successful adjustment.}.\\

$V^A$ or {\bf ``adjusted FPC''} is the  $V^B \cup V^D$ with cross covariances (correlations)
added with sufficient accuracy to obtain combined covariance matrix as
positive
semidefinite matrix. \\

$V^R$ or {\bf ``recommended FPC''} is the $V^A$ but rounded by NIST to be compactly
presented in their publications and as recommended data for science and
technology by CODATA.\\

All data sets $V^I$ defined above have the same pair of structures:
$$V^I = \{Average(V^I), Covariance(V^I)\}$$
or
$$V^I = \{Average(V^I), Uncertainty(U^I), Correlator(C^I)\} .$$

Let us call the internal calculational accuracy of numerical
presentation of all components of the $V^B$ obtained from the
adjustment procedures as {\bf etalon accuracy}.


\subsection{Correctness and Self-consistency}
If the adjustment of the constants belonging to $V^B$ is
successful then we have positive definite covariance (correlation)
matrix presented with an etalon accuracy, as well as the vector of
average values.

We say that the $V^D$, $V^A$ are correct if their covariance
(correlation) matrices are positive semi-definite.  In other words,
we have sufficient internal calculation accuracy to obtain  correct results.\\

\noindent
We say that the $V^R$ is correct and self-consistent if one of two possibilities
is true:

1) $V^R \equiv V^A$ or

2) For any
subset  $v(V^R) \subset Average(V^R)$ for which corresponding covariance submatrix $Cov(v(V^R))$ is positive definite we have

$$[v(V^R)-v(V^A)]_i \cdot[Cov(v(V^A))]_{ij}^{-1}\cdot[v(V^R)-v(V^A)]_j \le 1 $$
or
$$[v(V^R)-v(V^A)]_i\cdot [Cov(v(V^R))]_{ij}^{-1}\cdot[v(V^R)-v(V^A)]_j \le 1 .$$
These conditions guarantee the {\bf self-consistency} of the $V^R$, e.g.
that the rounded and unrounded scatter ellipsoids are well intersected and unrounded and rounded subvectors belong to that intersection.

\subsection{Reliability}
We will say that the next release $V^R_{YY}$ is reliable if it is correct, selfconsistent,
and if any subvector $v(V^R_{YY})$ with positive definite covariance is ended in the point inside the
scatter ellipsoid for the corresponding subvector of the
previous release. For example, for the 1998 and 2002 releases these conditions
will read
$$[v(V^R_{02}) - v(V^A_{98})]_i\cdot[Cov(v(V^A_{98}))]_{ij}^{-1}\cdot[v(V^R_{02})
-v(V^A_{98})]_j \le 1 .$$
The reliability indicator is constructed
with an assumption that the relative time variation of the
fundamental constants during two successive sessions of the
adjustments are negligible compared with the average relative
standard deviation of the constants.

\subsection{Availability}
Next important quality indicator we propose is the availability of
all data on FPC (average values, uncertainties, correlations) in
computer readable forms with as maximal as possible completeness
and accuracy of numerical data. The importance of the availability
is hard to overestimate in the era of the Web communications and
Web and GRID computations\footnote{ To taste the importance of the
availability requirement we will recommend reader to try to check
our calculations presented in the motivation section, including
the correctness of data extraction from NIST publications and
site.}.

It turns out that NIST and CODATA, in spite of the nicely
organized affiliation web cites offer the current and archived
data on the FPC in the hopeless obsolete manner, as it will be
shown in the sections to follow.

\subsection{Traceability}
The traceability in the context of usage the recommended FPC is
the access to all input experimental and theoretical material used
in the adjustment as well as detailed descriptions of the used
procedures needed to reproduce the adjustment independently in
case of any suspicions on the misprints in the database,
ideological or software bugs.\\

\section{Safety rounding off the correlated quantities}

Here we derive a simple sufficient estimates on the accuracy of a safely
independent rounding off the
average values  $V_i$, uncertainties $U_i$, correlations $C_{ij}$ obtained in jointly measurement or estimation procedures with sufficient etalon
accuracy.

Let  $(V_i, \, U_i, \,C_{ij})$, $i,j = 1, \dots, n$ be the  aggregate of
 $n$  jointly measured or estimated physical quantities, where
numerical parts of  $V_i, \, U_i$ are the real
numerical vectors, $U_i > 0$, $C_{ij}$ is the real, symmetric, and positive definite matrix with matrix elements bounded as follows:
$$C_{ii}= 1 \quad {\rm for\, all} \quad i=1,\dots,n \quad {\rm and}
\quad  |C_{i \ne j}| < 1.0.$$ Suppose that for some reason we need to
store and exchange numerical data on this aggregate rounded to some
accuracy $A$ that is lower than the etalon one.

\noindent
Let $R_{ij}$ be the ``rounder'' matrix, such that if it is added
to the matrix $C_{ij}$, the obtained matrix $C^R_{ij} = C_{ij} + R_{ij}$
will be real, symmetric, positive definite and all
$|C^R_{i\ne j}| < 1$ are decimal numbers with $A$ digits to the right of
 the decimal point.

It is easy to see that matrix $R_{ij}$ should have the following properties:
$$R_{ii}= 0 \quad {\rm for\, all} \quad i=1,\dots,n \quad {\rm and}
\quad  |R_{i \ne j}| \le  5.0 \times 10^{-A-1}.$$

Let further $c_1 \le \dots \le c_n$,
$\rho_1 \le \dots \le \rho_n$, and
$c^R_1 \le \dots \le c^R_n$ be the ordered sets  of eigenvalues
of the matrices  $C_{ij}$,  $R_{ij}$, and  $C^R_{ij}$ correspondingly.
Then from the Weil's theorem for any $l = 1,\dots,n$ we have
the following  inequalities \cite{HornJohnson},\cite{Wilkinson}:

$$c_l+\rho_1 \le c^R_l \le c_l + \rho_n.$$

>From the Gershgorin's theorem on the distributions of the eigenvalues of
the Hermitian matrices \cite{HornJohnson}  it follows that
$$\rho_1 \ge -(n-1)\cdot 5\cdot 10^{-(A+1)} =
-{\frac{(n-1)} {2}}\cdot10^{-A}$$
 and hence to have the matrix $C^R_{ij}$  as positive semi
definite matrix it is sufficient to demand
$$0 \le c_1 - {\frac {(n-1)} {2}}\cdot10^{-A} \le c^R_1.$$

>From the left inequality we have the final estimate for the
threshold accuracy index for safely uniform independent rounding of the
positive definite correlation matrix $C_{ij}$ with minimal
eigenvalue $c_1=\lambda^C_{min}$
{\boldmath \large
\begin{equation}
{ A \ge A_{C}^{th} = \left \lceil \log_{10}\left({\frac{n-1}
{2 \cdot \lambda^C_{min}}}\right ) \right \rceil }.
\label{K_SIR}
\end{equation}
}

\noindent
{\bf NOTE.} According to the Weil's theorem  any uniform rounding
 the off-diagonal matrix elements of the positive semi-definite correlation
 (covariance) matrix is forbidden.

 Indeed, as rounder
matrix is traceless Hermitian matrix, it obliged to have the
negative minimal eigenvalue. Furthermore from the left inequality
of the Weil's theorem statement it follows that any rounding
could lead to the matrix with negative minimal eigenvalue.

Now let us clarify to what accuracy we may round off the $V_i$ and
$U_i$ in the decimal presentations. Let $R^V_i$ be the such ``rounding vector'' that the obtained rounded vector  $V^R_i = V_i - R^V_i$ is still in the etalon
scatter ellipsoid. Then from the condition (\ref{corellipsoid}) for the components of the rounding vector we will have
\begin{equation}
\sum_{ij}{\frac {R^V_i} {U_i}}
\cdot [C^{-1}]_{ij}\cdot {\frac {R^V_j} {U_j}} < 1.
\label{valellipsoid}
\end{equation}
In the eigenbasis of the etalon correlator $C_{ij}$ the expression (\ref{valellipsoid})
can be transformed to
\begin{equation}
\sum_{ij}\sum_{mn}{\frac {R^V_i} {U_i}}\cdot[L^{-1}]_{im}
\cdot {\frac {\delta_{mn}} {\lambda_m}}\cdot [L]_{nj} \cdot {\frac {R^V_j} {U_j}} < 1,
\label{valell1}
\end{equation}
where $L$ is a rotation matrix. As we try to find the sufficient
condition for rounding vector components it is enough to demand
the validity of (\ref{valell1}) for all correlator eigenvalues
replaced with minimal one. Then the inequality (\ref{valell1})
will become

\begin{equation}
\sum_{i}\left ( \frac {R^V_i} {U_i}\right )^2 < \lambda^C_{min}.
\label{valell2}
\end{equation}

Inequality (\ref{valell2}) means  that we can round components
independently only inside the maximal hypercube imbeded into
scatter ellipsoid:
\begin{equation}
 {\frac {|R^V_i|} {U_i}}  < \sqrt{\frac {\lambda^C_{min}} {n}}.
 \label{hypercube}
\end{equation}
To obtain the accuracy $A^V_i$ for the $i$-th component that will be sufficient to guarantee that
the end of the vector $V^R_i$ belongs to the interior of the etalon scatter ellipsoid it is sufficient to have
$$|R^V_i| [unit_i] \le 5 \cdot 10^{-(A^V_i + 1)} [unit_i] .$$

\noindent
>From this bound it follows that to have the rounded vector of average
values pointing to the interior of the etalon scatter ellipsoid one should save
\begin{equation}
A_i \ge A^V_i = \left \lceil {\frac 1  2} \log_{10}\left({\frac {n} {4 \cdot
 \lambda^C_{min} \cdot (U_i/[unit_i])^2}} \right ) \right \rceil
\label{K_VIR}
\end{equation}
 digits to the right of the decimal point. \\



Now let us turn to the rounding of the uncertainties $U_i$. It is
the common practice to present the average values and
uncertainties with the same accuracy $A^V_i = A^U_i$.  With this
rule let us rewrite inequality (\ref{hypercube}) in the form

$$\log_{10}(U_i) \ge \log_{10}\left( {\frac 1 2} \sqrt{\frac {n}
{ \lambda^c_{min}}} \ \right)
-A^U_i .$$

 Taking into account the equality\footnote{This equality is valid for real numbers
only. For the integer number that treated as the numbers with
infinite precision it is not valid.}

 $$\lfloor \log_{10}(U_i) \rfloor +1 = P^U_i - A^U_i,$$ where $P^U_i$ is the precision of the $U_i$ we will obtain

\begin{equation} \label{K-eq}
P^U_i \ge \left \lceil {\frac 1 2}
\log_{10}\left({\frac {n} {4 \cdot
 \lambda^C_{min}}} \right ) \right \rceil .
\end{equation}

One can see that right part of the inequality does not depend on
index $i$, so we can introduce $P^U$ which is the same for every
$i$: $$P^U=P^U_i .$$

The equation (\ref{K-eq}) give the minimal precision that should not
be reduced if we adopt the rule that accuracy of the uncertainties
should be equal to the accuracy of the average
values. \\


\noindent
In summary: we have obtained $n+1$ reference numbers $A^{th}_C$ and $A^V_i$ defining the levels with safety independent rounding off the decimal numerical presentation of the  interrelated random quantities: average values, their
uncertainties, and correlations.\\

\noindent
 Having these numbers the strategy for the safety independent rounding can be as follows: \\

\noindent {\large \it In self-consistent numerical presentation of
interrelated random quantities $( V_i,\, U_i, \, C_{ij})$ in
decimal real numbers the average values  $V_i$ and the
uncertainties $U_i$ should have at least $A^V_i$ digits to the right of the
decimal point and the correlation coefficients $C_{i \ne j}$
should have at least $ A^{th}_C$ digits to the right of the decimal
point.} \\

\section{ Do the CODATA 2002 recommended FPC  meet the high level
quality requirements?}

In this section we present some further evidences of violations of the
above high level requirements in the recent releases of the CODATA
recommended values of the FPC.

\subsection{Correctness \& Selfconsistency}
In motivation section we already presented the evidences that the
CODATA data on correlations are incorrect. Here we present an
evidence that the average values of the recommended FPC are also
questionable, because of over-rounding can easily move them out of
the etalon scatter ellipsoid. To check this the whole adjustment
process should be repeated with the ``etalon accuracy''.


 It turned out that  we managed to collect enough
 amount of data from the NIST publications to reproduce all steps of the evaluation and adjustment of the basic
 set of constants
 \cite{ASSite} only for the 1998 release. We had obtained the ``correct set of the basic constants''  using methods
 described by NIST experts \cite{CODATA:98}\footnote{As the correlation matrix of the uncertainties in the input
 experimental data is not a positive
 definite matrix there (supposedly by overrounding for publication), we were forced to ``un-round''
  several matrix elements to have positive definite weight matrix in the least squares method of adjustment.}
and then calculated the threshold accuracies for the elements of
the correlation matrix, the averages and the uncertainties. The
results are as follows:
$$\lambda_{C,min}\approx7.58\cdot10^{-7} ,$$
$$A^{th}_C=8 \:{\rm (versus}\:A_C^{CODATA}=3{\rm)},$$
$$P^U=4 \:{\rm (versus}\:P^{U, CODATA}=2{\rm)}.$$
One can see that the CODATA data suffers the loss of
 self-consistency of the released data due to unjustified
 over-rounding of their results.

Having the data on the FPC in the ``etalon accuracy'' we are
able to show that the obtained estimates for the threshold
rounding indices are indeed close to the real situation and should
be used as regulators for the correctness of the rounding. To show
that the rounding procedure can move the end of the
vector-of-constants out of the etalon scatter ellipsoid we will use the sample
of constants that was mentioned in \cite{Uzan:2002vq} as the
candidates to trace the large-scale space-time variability of
their dimensionless combinations:
\begin{table}[h]
\label{ihepdata}
\begin{center}
\vspace*{-3mm}
\caption{Selected basic and derived constants from the IHEP adjustment based
on the NIST 1998 input data.}
\vspace*{1mm}
\begin{tabular}{|l||ll|} \hline
Symbol[units] & ~~~~~Average value & Uncertainty \\ \hline \hline
& & \\[-4mm]
 $\quad h \quad \quad [{\rm J} \, {\rm s}]$  &
$6.62606875610000 \times {10}^{-34}$ & $5.2200000 \times {10}^{-41}$ \\
$\quad m_e \, \quad [{\rm kg}]$
& $9.10938187491360 \times {10}^{-31}$ & $7.2057063 \times {10}^{-38}$\\
$\quad m_p  \, \quad [{\rm kg}]$ & $1.67262158291420 \times {10}^{-27}$ &
$1.3235274 \times {10}^{-34}$\\
$\quad m_n \, \quad [{\rm kg}]$ & $1.67492715608612 \times {10}^{-27}$ & $1.3253602 \times {10}^{-34}$\\ $\quad e \, \, \quad \quad [{\rm C}]$  & $1.60217646198672 \times {10}^{-19}$ & $6.3181739 \times {10}^{-27}$\\
\hline
\end{tabular}
\end{center}
\end{table}

\noindent
The corresponding  correlation matrix of their uncertainties in the ``etalon
accuracy''~\footnote{The Plank constant is the basic one, the other selected
are derived constants. In calculating the corresponding correlation matrix we use the minimal possible accuracy that give us the positive definite
correlation matrix.}
\begin{center}
\begin{tabular}{|c|c|c|c|c|c|}
\hline $Cor$& $h$ & $m_e$ & $m_p$ & $m_n$ & $e$\\ \hline

$h$ & 1.000000000 & 0.9957673366 & 0.9954294463 & 0.9954234131& 0.9989373297
\\

$m_e$ & 0.9957673366 & 1.000000000 & 0.9996433868 & 0.9996224521 &
0.9904731204\\

$m_p$ & 0.9954294463 & 0.9996433868 & 1.000000000 & 0.9999732991 &
0.9901455374\\

$m_n$ & 0.9954234131 & 0.9996224521 & 0.9999732991 & 1.000000000 &
0.9901469965\\

$e$ & 0.9989373297 & 0.9904731204 & 0.9901455374 & 0.9901469965 & 1.000000000\\
\hline
\end{tabular}
\end{center}

\noindent is the positive definite matrix with eigenvalues as follows: $$\{4.98223,\,
0.0172451,\, 0.000495716,\,
    0.0000263673,\, 6.47023 \times 10^{-10}\}.$$
Corresponding $A^{th}_C=10$ and it is close enough to our minimal accuracy, the rounding
of the above correlator to $8$ digits will make the matrix non-positive definite.

Now we will round average values of the constants to have accuracy below the allowed
thresholds $A^V_i$. In the Table 3
 we present the values of the
differences  $\langle V_i \rangle - V^r_i$ between calculated average values of the
selected constants with the etalon accuracy and the rounded step-by-step values to show
that after the predicted moment the end point of the rounded vector will be moved out of
the etalon scatter ellipsoid for many standards $R(V^r,\langle V \rangle)$.

\begin{table}[h]
\begin{center}
\vspace*{-3mm}
\caption{Evolution of the ``distance'' of the end point of rounded
vector from the etalon scatter ellipsoid expressed in number of
standard deviations squared with rounding off the
vector components $R_i^V$ in steps.} \vspace*{2mm}
\begin{tabular}{|c||c|c|c|c|c||l|}
\hline
 & & & & & & \\[-4mm]
 ${\rm Step}$ & $ \phantom{-}h $ [J s] & $ \phantom{-}m_e$
 [kg] & $ \phantom{-}m_p $ [kg] & $ \phantom{-}m_n $ [kg] & $ \phantom{-}e $ [C] & $ R^2(V^r,\langle V \rangle) $ \\ \hline
$ 9 $ & $\phantom{-}4.39$E-42 & $\phantom{-}2.51$E-39 & $ \phantom{-}1.71$E-35& $\phantom{-}4.39$E-35& $ -1.99$E-28 & $3.9$E+06 \\
$ 8 $ & $\phantom{-}3.90$E-43 & $ -4.91$E-40 & $ -2.25$E-36 &
$\phantom{-}3.91$E-36 & $ \phantom{-}1.40$E-30 & $4.1$E+04  \\
$ 7 $ & $-9.52$E-45  & $ \phantom{-}8.92$E-42 & $\phantom{-}8.49$E-38&
$-8.70$E-38 & $ \phantom{-}1.40$E-30 & $ 61. $ \\
 6  &  \phantom{-}4.79 E-46 & -1.08E-42 &  -1.51E-38 &
 \phantom{-}1.30E-38 & \phantom{-}4.02E-31 &  0.36  \\
 5  &  \phantom{-}4.79E-46 &  -7.59E-44&  \phantom{-}4.90E-39&
 \phantom{-}3.03E-39 &  \phantom{-}2.29E-33 &  0.038 \\
4  &  -2.12E-47 &  \phantom{-}2.41E-44 &  -1.01E-40 &
\phantom{-}3.16E-41 &  \phantom{-}2.29E-33 &  0.00026  \\
 3  &  -1.23E-48 &  \phantom{-}4.14E-45 &  -1.38E-42 &
 \phantom{-}3.16E-41 & \phantom{-}2.89E-34 &  2.5E-06 \\
 2  &  -2.32E-49 & \phantom{-}1.37E-46 &  -1.38E-42 &
\phantom{-}1.62E-42 & -1.09E-35 &  4.5E-09 \\
 1  &  -3.16E-50 & \phantom{-}3.74E-47 &  -3.84E-43 &
 -3.64E-43 & -9.03E-37 &  2.7E-09 \\
 0  &  -1.58E-51 & -2.57E-48 &  \phantom{-}1.65E-44 &
 \phantom{-}3.61E-44 & \phantom{-}9.69E-38 & 6.1E-14 \\
\hline
 \hline
  & & & & & & \\[-4mm]
 $ A^V_i $ & $ 45 $ & $ 42 $ & $ 39 $ & $ 39 $ & $ 31 $ & $  $ \\
\hline
\end{tabular}
\end{center}
\label{roundihep}
\end{table}

We see that our indices proposed as the sufficient number of
digits for the safety rounding are indeed close to the reality.
They can and should be used to the quality control of the random
vectors obtained by statistical estimation procedures.

Another lesson from the comparisons presented above is that the problem of
the correct rounding off the FPC triad $(V_i,\, U_i, C_{ij})$ is the very
important problem in the task of tracing the space-time variability of the
FPC as the improper rounding will mimic the evolution of constants.

The third lesson is that the CODATA recommended values of the FPC are
highly questionable as we have convinced that the correlation matrices
were corrupted by the unjustified rounding.

\subsection{Reliability}
As it was mentioned in the descriptions of the high level quality
requirements, it is natural to suppose that the next iteration of
the adjustment will give constants more accurate and more
selfconsistent than the previous adjustments.

Let us look for the time evolution of the estimates of one of the most important physical
constant --- the Planks constant $h$ from the time of discovery up to the 2002 estimate.
The historical perspective of the Plank constant estimates one can find in
\cite{VNLVVE:2000}.
\newpage


\begin{figure}[h]
\centering\noindent
\vspace*{3mm}
\caption{\normalsize  Plank Constant: 1969--2002. Error band show that
the adjustment procedures produce estimates that still are far from been stable,
though the amplitude of variation is reduced in the last two releases.}
\end{figure}
%
\vspace*{7mm}
This  ``small-scale time variability'' of the Plank constant
estimates we attribute to the possible presence of the hidden (not
estimated) systematic error introduced or missed by the adjustment
procedures. It should be noted that systematist have to use
contradictory input data which impossible to refine at the time of
adjustment sessions~\footnote{See discussion of this issue in the
subsection:
 ``A. Comparison of 1998 and 1986 CODATA recommended values'' of the summary
 of the 1998 review (\cite{CODATA:98}, pages 459-461).}.

The ``evidence'' of the possible stabilization (see Fig.~2
is very preliminary and should
be tested for the other constants simultaneously by tracing the variation
of the hodograph of the ``vector of basic constants'' as it is outlined in the reliability requirement. Unfortunately it is not possible now because of the corrupted data on correlations in the releases. The conclusion based on
the reliability indicator is that the CODATA recommended values cannot
be used in searches of the possible large scale space-time variations
of the FPC.

%
\newpage
\begin{figure}[h]
\label{stabilization} \centering\noindent
 \caption{\normalsize
 Plank Constant: 1986--2002. Evidence for the possible
 stabilization.}
\end{figure}
\vspace*{2mm}

\subsection{Availability}

The web access to the data on FPC offered by NIST \& CODATA in the
last release (V.4.0) is greatly improved. Now we have easy access
to all data on average values and their uncertainties just
copy the file in the ASCII format. But unfortunately in the
released list the values of 7 basic constants out of 29
participating in the adjustment process did not quoted. The values
of the other 28 important parameters (possible corrections to the
theoretical expressions) for the whole adjustment procedure are
omitted. They even did not discussed in the publications on the
1998 release.

As it was discussed in the previous sections, the ignorance of the
correlations is inadmissable in the high precision physics applications.
But the access to the recommended correlation coefficients remains to be
the ``misanthropic'' one.
 It is hard to get data for an operative calculations
with several constants simultaneously. There is no easy and safety
 way to get the complete data on the subsample of the triad $( V_i, U_i, C_{i,j})$ in a truly computer readable form.

To extract data on say 10 constants with the correlation matrix
one have to produce about 300 flip-flops between web-pages ``by
hands''.

\subsection{Traceability}

Traceability means that any release of the recommended FPC set
should be accompanied with full toolkit of the input data and
methods to give interested user possibility to perform all steps
of the adjustment process and to compare the results with the
recommended values.

Unfortunately materials attached to the
recommended FPC are not complete as it was stressed in the
discussions of the availability indicator. Additional example is
the incorrect presentation of the correlations of uncertainties in
the input experimental data of the 1998 release.

The data on input
correlations are presented only in the review on the paper
\cite{CODATA:98}  and the correlation matrix is non positive
definite there \cite{ASS-VVE:03}.

It should be noted also that in
the published documents related to the releases of FPC  there are
no discussions of the procedures used for rounding off the
correlated quantities.

\section{Summary}

Summarizing the above discussions and evidences we are forced to stress that
all high level quality requirements to the scientific information numerical
data resource: correctness, selfconsistency, availability, reliability, and
traceability are badly violated in at least the last three releases of the CODATA recommended values of the fundamental physical constants.

They could not be used as reference data to monitor the large scale
space-time variability of the fundamental physical constants and
moreover their usage in physics applications where the high
precision calculations are needed is highly questionable.

The positive outcome from our critical treatment of the quality aspects of
the central numerical scientific information resource are:
\begin{itemize}
\item the preliminary proposal for the safety rounding strategy in presentation the results of high precision computations of the physical observables;
\item the  proposal for the set of quality indicators
to certify scientific information resources for the safety usage in physics
applications;

\item the  proposal of the data structure and procedures for the complete and user friendly \href{http://sirius.ihep.su/~kuyanov/beta/PhysConst/FPC-1998-4.html} {Web-FPC}.
\end{itemize}

\enterdate{September 9, 2004.}

\outputnames
{В.В.Ежела, Ю.В.Куянов, В.Н.Ларин, А.С Сивер.}
{Непостоянство фундаментальных физических постоянных: вычислительный статус}
{2004--36}
{авторами}
{}
 {}{}
\outputdata
 {13.09.2004.}
 {1,75.}
 {1,4.}
 {160.}
 {304}
 {3649.}
 {}
\end{document}